\documentclass[aps,prb,twocolumn]{revtex4-2}

\usepackage{graphicx}
\usepackage{upgreek}
\usepackage{amsmath}
\usepackage{amssymb}
\usepackage{dcolumn}
\usepackage{chngpage}

\usepackage{CJK}

\usepackage{graphicx}% Include figure files
\usepackage{dcolumn}% Align table columns on decimal point
\usepackage{bm}% bold math
\usepackage{amssymb} % added by EJF for some math symbols
\usepackage{subfigure}% Include figure files
\usepackage{color}
\usepackage{epstopdf}
\usepackage{tabularx}

\begin{document}

\title{Emergence of the spin polarized domains in the kagome lattice Heisenberg antiferromagnet Zn-barlowite (Zn$_{0.95}$Cu$_{0.05}$)Cu$_{3}$(OD)$_{6}$FBr}% Force line breaks with \\

\author{Weishi Yuan$^{1,6}$}
\author{Jiaming Wang$^{1,6}$}
\author{Philip M. Singer$^{2}$}
\author{Rebecca W. Smaha$^{3,4}$}
\author{Jiajia Wen$^{3}$}
\author{Young S. Lee$^{3,5}$}
\author{Takashi Imai$^{1}$}%\email{imai@mcmaster.ca}
\email[Corresponding author: ]{imai@mcmaster.ca}

\affiliation{$^{1}$Department of Physics and Astronomy, McMaster University, Hamilton, Ontario, L8S 4M1, Canada}
\affiliation{$^{2}$Department of Chemical and Biomolecular Engineering, Rice University, 6100 Main St., Houston, TX 77005, USA}
\affiliation{$^{3}$Stanford Institute for Materials and Energy Sciences, SLAC National Accelerator Laboratory, Menlo Park, CA 94025, USA}
\affiliation{$^{4}$Department of Chemistry, Stanford University, Stanford, CA 94305, USA}
\affiliation{$^{5}$Department of Applied Physics, Stanford University, Stanford, CA 94305, USA}
\affiliation{$^{6}$These authors made equal contributions.}

%\affiliation{%


%\affiliation{%



\begin{abstract}
Kagome lattice Heisenberg antiferromagnets are known to be highly sensitive to perturbations caused by structural disorder.  NMR is a local probe ideally suited for investigating such disorder-induced effects, but in practice large distributions in the conventional one-dimensional NMR data make it difficult to distinguish the intrinsic behavior expected for pristine kagome quantum spin liquids from disorder induced effects.  Here we report the development of a two-dimensional NMR data acquisition scheme applied to Zn-barlowite (Zn$_{0.95}$Cu$_{0.05}$)Cu$_{3}$(OD)$_{6}$FBr kagome lattice, and successfully correlate the distribution of the low energy spin excitations with that of the local spin susceptibility.  We present evidence for the gradual growth of domains with a local spin polarization induced by 5\% Cu$^{2+}$ defect spins occupying the interlayer non-magnetic Zn$^{2+}$ sites.  These spin polarized domains account for $\sim60$\% of the sample volume at 2~K, where gapless excitations induced by interlayer defects dominate the low energy sector of spin excitations within the kagome planes.
\end{abstract}

%\begin{abstract}
%An article usually includes an abstract, a concise summary of the work
%covered at length in the main body of the article. 
%\begin{description}
%\end{description}
%\end{abstract}

\maketitle

%\tableofcontents

\noindent{\bf INTRODUCTION}\\

The quantum spin liquid (QSL) is a novel state of matter formed by entangled spin singlets, in which magnetic frustration effects prevent spins from undergoing magnetic long range order \cite{Balents2010,Norman2016,Zhou2017,Broholm2020,Imai2016}.  The last few decades have seen concerted efforts to identify model materials of the QSL.  However, many of them undergo magnetic long range order or spin freezing at low temperatures.  Copper hydroxyhalide materials herbertsmithite ZnCu$_{3}$(OH)$_{6}$Cl$_2$ ~\cite{Shores2005,Helton2007,Mendels2007,Rigol2007,Rigol2007_PRB,Imai2008,Olariu2008,Zorko2008,Freedman2010,Imai2011,Han2012,Fu2015,Sherman2016,Zorko2017,Khunita2020,Wang2021,Huang2021,Murayama2021}  and Zn-barlowite ZnCu$_{3}$(OH)$_{6}$FBr ~\cite{Feng2017,Smaha2020,Smaha2020_PRM,Tustain2020,Wei_2020,Fu:2021aa,Wang2021,Wang2022} are among the few exceptions without magnetic symmetry breaking even near absolute zero.  As shown in Fig.1, Cu$^{2+}$ ions  with electron spin-$1/2$ in Zn-barlowite form a corner-sharing triangular lattice known as the kagome lattice.  The nearest-neighbor Cu$^{2+}$-Cu$^{2+}$ antiferromagnetic super-exchange interaction is as large as $J \sim 160$~K~\cite{Helton2007,Smaha2020} and geometrically frustrated.  While earlier studies point toward realization of a proximate kagome QSL state in these materials~\cite{Shores2005,Helton2007,Mendels2007,Imai2008,Olariu2008,Imai2011,Han2012,Fu2015,Zorko2017,Khunita2020,Wang2021,Huang2021,Murayama2021,Feng2017,Tustain2020,Wang2021}, the structural disorder in these materials~\cite{Freedman2010,Smaha2020_PRM} complicates interpretation of the experimental results.  Accordingly, the exact nature of the kagome planes has been highly controversial, necessitating clear understanding of the influence of disorder on the kagome planes.

\begin{figure}
	\begin{center}
		\includegraphics[width=1\columnwidth]{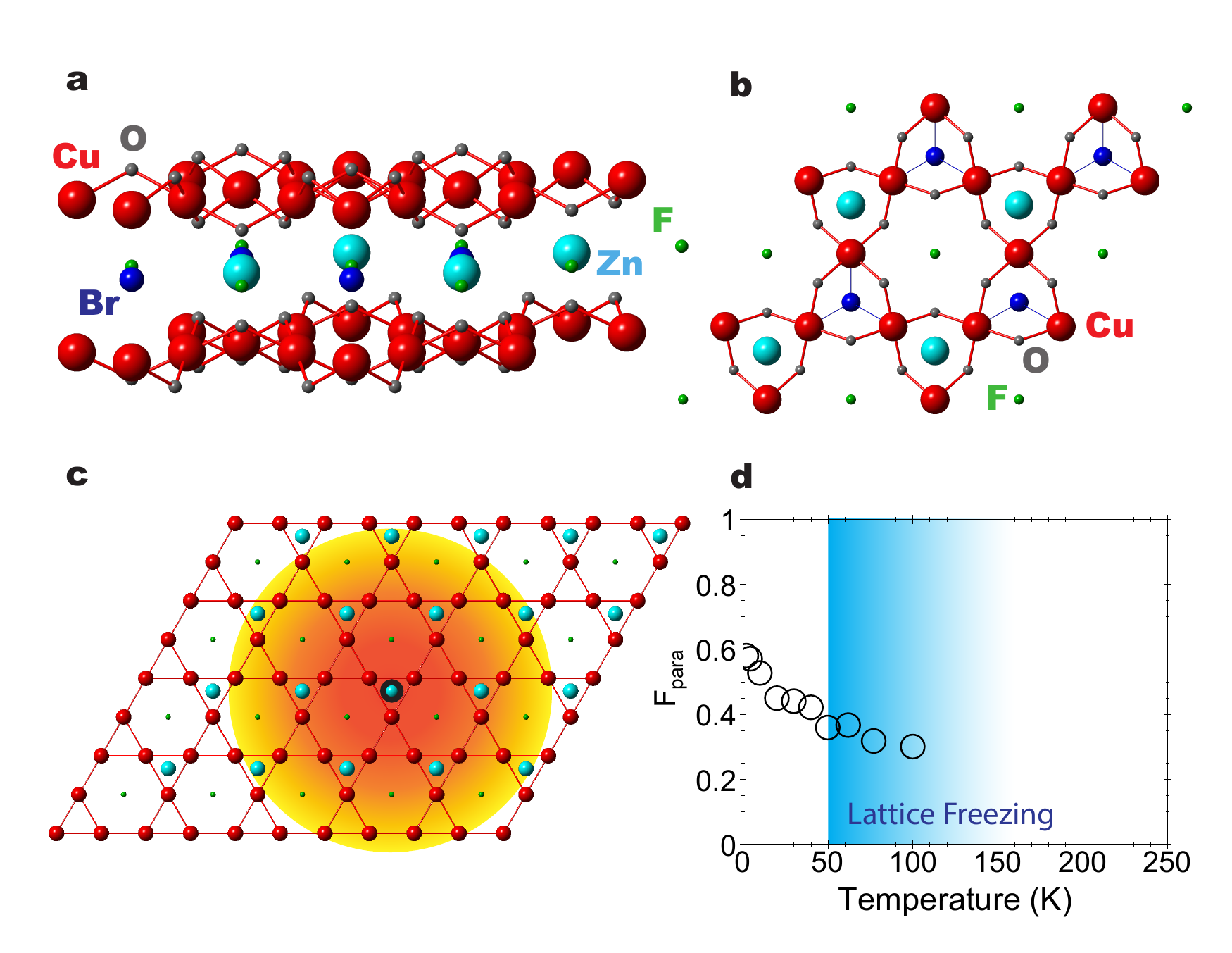}
		\caption{{\bf Kagome lattice}: {\bf a} The crystal structure of Zn-barlowite ``Zn$_{0.95}$" (Zn$_{0.95}$Cu$_{0.05}$)Cu$_{3}$(OD)$_{6}$FBr.  For clarity, D-sites attached to O-sites are not shown in any of the panels.  {\bf b} A c-axis view of the kagome plane of Zn$_{0.95}$, and the interlayer Zn (cyan), Br (blue), and F (green) sites above and below the kagome plane.  {\bf c} A wider field of view of the kagome plane in Zn$_{0.95}$, without D, O, and Br sites.  The Zn$^{2+}$ site surrounded by a black circle near the middle of the panel represents the Cu$^{2+}$ interlayer defect occupying one of the twenty Zn sites within this field of view with 5\% probability.  The circle with orange-to-yellow shading represents the spin polarized domains, which encompasses 12 of the 20 $^{19}$F sites in this field of view, corresponding to 60\% as observed at 2~K in panel {\bf d}.  All the kagome Cu sites in this domain have polarized spins or interact with them.  {\bf d} Temperature dependence of the fraction $F_\text{para}$ of the $^{19}$F sites belonging to the spin polarized domains estimated in this work.  Blue shading depicts the freezing of the lattice distortion starting at $\sim160$~K and completing at $\sim50$~K \cite{Wang2021}.}
		\label{fig:Fig1}
	\end{center}
\end{figure}

Recent work, including site-selective anomalous X-ray scattering experiments, established that the interlayer non-magnetic Zn$^{2+}$ sites are occupied by extra Cu$^{2+}$ defect spins with $\sim$15\% probability in herbertsmithite \cite{Freedman2010} and $\sim$5\% probability in the most Zn-rich Zn-barlowite, synthesized as a powder  \cite{Smaha2020_PRM}. Accordingly, their actual chemical formula is (Zn$_{0.85}$Cu$_{0.15}$)Cu$_{3}$(OH)$_{6}$Cl$_2$ for herbertsmithite and (Zn$_{0.95}$Cu$_{0.05}$)Cu$_{3}$(OH)$_{6}$FBr for Zn-barlowite. We will use an acronym Zn$_{0.95}$ to represent Zn-barlowite hereafter. One important difference between Zn-barlowite and herbertsmithite is in the location of the interlayer Cu$^{2+}$ defects: in herbertsmithite, they occupy the centered, octahedral Zn$^{2+}$ site, while in Zn-barlowite they occupy an off-center, trigonal prismatic site that is distinct from the octahedral Zn$^{2+}$ site~\cite{Smaha2020,Smaha2020_PRM}.  

The extra interlayer Cu$^{2+}$ defect spins in both of these materials could directly contribute to the magnetic response of the sample, such as enhanced bulk spin susceptibility $\chi_\text{bulk}$ at low temperatures.  In addition, each interlayer defect spin  interacts with three nearest neighbor (nn) kagome Cu$^{2+}$ sites located in each of the two adjacent kagome planes.  Furthermore, the structural distortion in the vicinity of the interlayer defects \cite{Imai2008,Fu2015,Wang2021,Wang2022} might enhance local spin susceptibility $\chi_\text{local}$ at low temperatures through Dzyaloshinskii-Moriya interaction \cite{Rigol2007,Rigol2007_PRB,Zorko2008,Zorko2017} and/or randomness introduced to the exchange interaction \cite{Shimokawa2015,Kimchi2018,Kawamura2019}.  On the other hand, the upper bound of the concentration of the non-magnetic Zn$^{2+}$  inter-site defects diluting the magnetic Cu$^{2+}$ sites within the kagome planes is $\sim1$\% in both of these materials \cite{Freedman2010,Smaha2020_PRM}.  This contradicts with the persisting claim that as much as $\sim$5\% of kagome Cu sites may be diluted by non-magnetic inter-site Zn$^{2+}$ defects, which cause local spin singlets accompanied by oscillatory spin density induced in their vicinity \cite{Olariu2008,Khunita2020}.  

Our earlier NMR measurements in the isotope-enriched single crystal samples of herbertsmithite successfully separated the main NMR peaks from those associated with the nn and the next nearest neighbor (nnn) $^{2}$D (nuclear spin $I=1$) \cite{Imai2011} and $^{17}$O (nuclear spin $I=5/2$) \cite{Fu2015} sites of the 15\% interlayer Cu$^{2+}$ defect spins.  We demonstrated that the Knight shift of the main NMR peak arising from the sites far away from the interlayer defects decreases toward zero.  The implication of the diminishing local spin susceptibility $\chi_\text{local}$ far from the defects has been the subject of an intense debate, {\it i.e.} whether the proximate kagome QSL realized in herbertsmithite and Zn$_{0.95}$ is gapped or gapless \cite{Imai2008,Olariu2008,Imai2011,Fu2015,Feng2017,Khunita2020,Wang2021}.  On the other hand, the NMR Knight shift at the nn sites of the interlayer defects in herbertsmithite is negative and its magnitude grows with decreasing temperature roughly in proportion to the bulk averaged spin susceptibility $\chi_\text{bulk}$ \cite{Imai2011,Fu2015}.  This means that the interlayer defect moments, which are polarized by the external magnetic field $B_\text{ext}$ applied to conduct NMR measurements, induce a negative hyperfine magnetic field $B_\text{hyp}$ pointing the opposite direction from $B_\text{ext}$.  In principle, one can probe the spatial extent of the influence of interlayer Cu$^{2+}$ defect spins on the adjacent kagome planes by investigating the nnn and further $^{2}$D  and $^{17}$O sites.  But we were unable to resolve these peaks at low temperatures, because the splitting between the multiple quadrupole-split NMR peaks is obscured by extreme magnetic line broadening with both positive as well as negative $B_\text{hyp}$.  

Very recently, we used inverse Laplace transform (ILT) T$_1$ analysis technique \cite{Song2002,Venkataramanan2002,Mitchell2012,Singer2018,Singer2020,Arsenault2020,Takahashi2019,Wang2021,WangPRB2021} to successfully deduce the density distribution function $P(1/T_{1})$ of the nuclear spin-lattice relaxation rate $1/T_{1}$ in herbertsmithite and Zn$_{0.95}$ \cite{Wang2021}.  The integral of $P(1/T_{1})$ is normalized to 1, and $P(1/T_{1})$ represents the probability for the nuclear spin to relax with a given value of $1/T_{1}$.  In general, $1/T_1$ probes low energy spin excitations in the form of the wave vector $\bf{q}$ integral of the dynamical electron spin susceptibility at the NMR frequency $\omega~(= 2\pi f)$,
\begin{equation}
\frac{1}{T_{1}} = \frac{\gamma_{n}^{2}k_{B}T}{\mu_{B}^2\hbar^{2}} \sum_{\bf{q}} |A_{\bf{q}}|^{2} \frac{\chi''({\bf q}, \omega)}{\omega},
\label{eq:Stretch}
\end{equation}
where $A_{\bf{q}}$ is the $\bf{q}$ dependent hyperfine form factor, and $\chi''({\bf q},\omega)$ is the imaginary part of the dynamical electron spin susceptibility (i.e. spin fluctuations at the resonance frequency $\omega$).

It turned out that the $^{63}$Cu nuclear spin-lattice relaxation rate $1/T_1$ in Zn$_{0.95}$ develops a bimodal distribution below $\sim30$~K due to the gradual emergence of spin singlets with inhomogeneous gap $\Delta$, which has a large distribution ranging from a few K to as high as $\Delta \sim30$~K \cite{Wang2021}.  These Cu sites involved in spin singlet formation exhibit diminishing $1/T_1$ with decreasing temperature, and the absolute upper bound of their volume fraction is $F_\text{singlet}\sim50$~\%.  On the other hand, the remaining Cu sites with the fraction of $1-F_\text{singlet}\sim50$~\% or greater do not participate in spin singlet formation, and remain correlated but paramagnetic.  These paramagnetic Cu spins exhibit a large and roughly constant $1/T_{1}\sim T \chi'' \simeq 10^{3}$~s$^{-1}$, which is typical for gapless excitations of paramagnetic spins coupled by super-exchange interactions, but their origin was not clear.

The aim of this study is to probe the origin, nature, and fraction of these paramagnetic Cu sites based on $^{19}$F NMR study of Zn$_{0.95}$.  $^{19}$F has a nuclear spin $I=1/2$, lacks nuclear quadrupole moment, and hence $^{19}$F NMR is immune from nuclear quadrupole interaction that splits and complicates $^{2}$D  and $^{17}$O NMR lineshapes.  $^{19}$F NMR therefore provides a unique avenue to probe the influence of interlayer defects on the kagome planes.  Nonetheless, earlier $^{19}$F NMR studies based on the conventional one-dimensional NMR approach \cite{Feng2017,Wang2021} did not provide a clear cut picture on the nature of defects in Zn$_{0.95}$, because the line broadening obscured the distinction between the intrinsic and defect induced phenomena. 

In this paper, we report the development of a two-dimensional NMR data acquisition scheme and its application to $^{19}$F NMR measurements in  Zn$_{0.95}$.   Instead of separately measuring the position by position variations of $\chi_\text{local}$ and low energy spin excitations from the distributions of the Knight shift and $1/T_1$, respectively, we generate a two dimensional correlation map $C(f; 1/T_{1})$ between them.  In essence, this procedure allows us to count the number of $^{19}$F sites that have certain values of the Knight shift $K$ {\it and} $1/T_1$ (i.e. $\chi_\text{local}$ {\it and} local spin excitations).  We demonstrate that interlayer Cu$^{2+}$ defect spins induce distinct spin polarized domains at low temperatures, in which $^{19}$F sites exhibit similar behavior as in antiferromagnetic barlowite Cu$_4$(OH)$_6$FBr  with N\'eel temperature $T_{N}=15$~K (abbreviated as Cu$_{4}$ hereafter) \cite{Han2014,Jeschke2015,Liu2015,Han2016_Barlowite,Guterding2016,Ranjith2018,Tustain2018,Smaha2020}.  The $^{19}$F sites in these domains of Zn$_{0.95}$ detect gapless spin excitations and enhanced $\chi_\text{local}$ induced by interlayer Cu$^{2+}$ defects spins.  The volume fraction of these domains reaches as much as $F_\text{para}\sim$60\% at 2~K despite the low 5\% concentration of the interlayer defects in Zn$_{0.95}$, indicating the gradual spatial growth of these domains at low temperatures.

The rest of this article is organized as follows.  We will begin the next RESULTS section with a presentation of the  one-dimensional NMR data.  We will explain how much one can learn from such conventional NMR data with the aid of ILT, and also their limitations.  The following sub-section will develop the two-dimensional NMR data acquisition scheme, and discuss how one can de-convolute the $^{19}$F NMR peak into two components with different magnitude of $1/T_1$.  In the DISCUSSION section, we will compare Zn$_{0.95}$ and Cu$_{4}$, and discuss the implications of our findings.\\

\noindent{\bf RESULTS}\\

\noindent{\bf Conventional one-dimensional NMR data and their limitations}\\

\indent We first summarize the conventional one dimensional $^{19}$F NMR results, and their limitations in elucidating the influence of defects.  In Fig.2a, we present the representative $^{19}$F NMR lineshapes observed for a deuterated, polycrystalline Zn$_{0.95}$ sample of (Zn$_{0.95}$Cu$_{0.05}$)Cu$_{3}$(OD)$_{6}$FBr in an external magnetic field of $B_\text{ext}=2.4$~T.  The full width at half maximum (FWHM) of the lineshape is as narrow as 26~KHz at 250~K owing to the small nuclear dipole moment of $^{2}$D.  Upon cooling, the peak frequency $^{19}f_\text{peak}$ of the lineshape slightly increases down to $\sim20$~K, then decreases toward the bare resonance frequency $^{19}f_\text{0} = ^{19}\gamma_{n} B_\text{ext} \simeq 96.02$~MHz marked by the vertical dashed line.  $^{19}f_\text{0}$ is the resonance frequency expected in the absence of hyperfine interactions with electron spins in the lattice, and the nuclear gyromagnetic ratio of $^{19}$F nucleus is $^{19}\gamma_{n}/2\pi = 40.05$~MHz/Tesla.  We can summarize the change of $^{19}f_\text{peak}$ using the NMR Knight shift $^{19}K_\text{peak}$ defined as 
\begin{equation}
^{19}K_\text{peak} = \frac{^{19}f_\text{peak} - ^{19}f_\text{0}}{^{19}f_\text{0}}
=\sum_{i=1}^{12} \frac{A_\text{hf}}{N_{A}\mu_{B}}~\chi^{i}_\text{local} + ^{19}K_\text{chem},
\label{eq:Stretch}
\end{equation}
where the summation goes over the twelve nn Cu sites of $^{19}$F nucleus with local spin susceptibility $\chi^{i}_\text{local}$, $N_{A}$ is Avogadro's number, $A_\text{hf}$ is the positive hyperfine coupling with these nn Cu sites \cite{Feng2017}, and $^{19}K_\text{chem}$($\simeq 0.01$\%) is a very small and temperature independent chemical shift \cite{Feng2017}.

\begin{figure}
	\begin{center}
		\includegraphics[width=1\columnwidth]{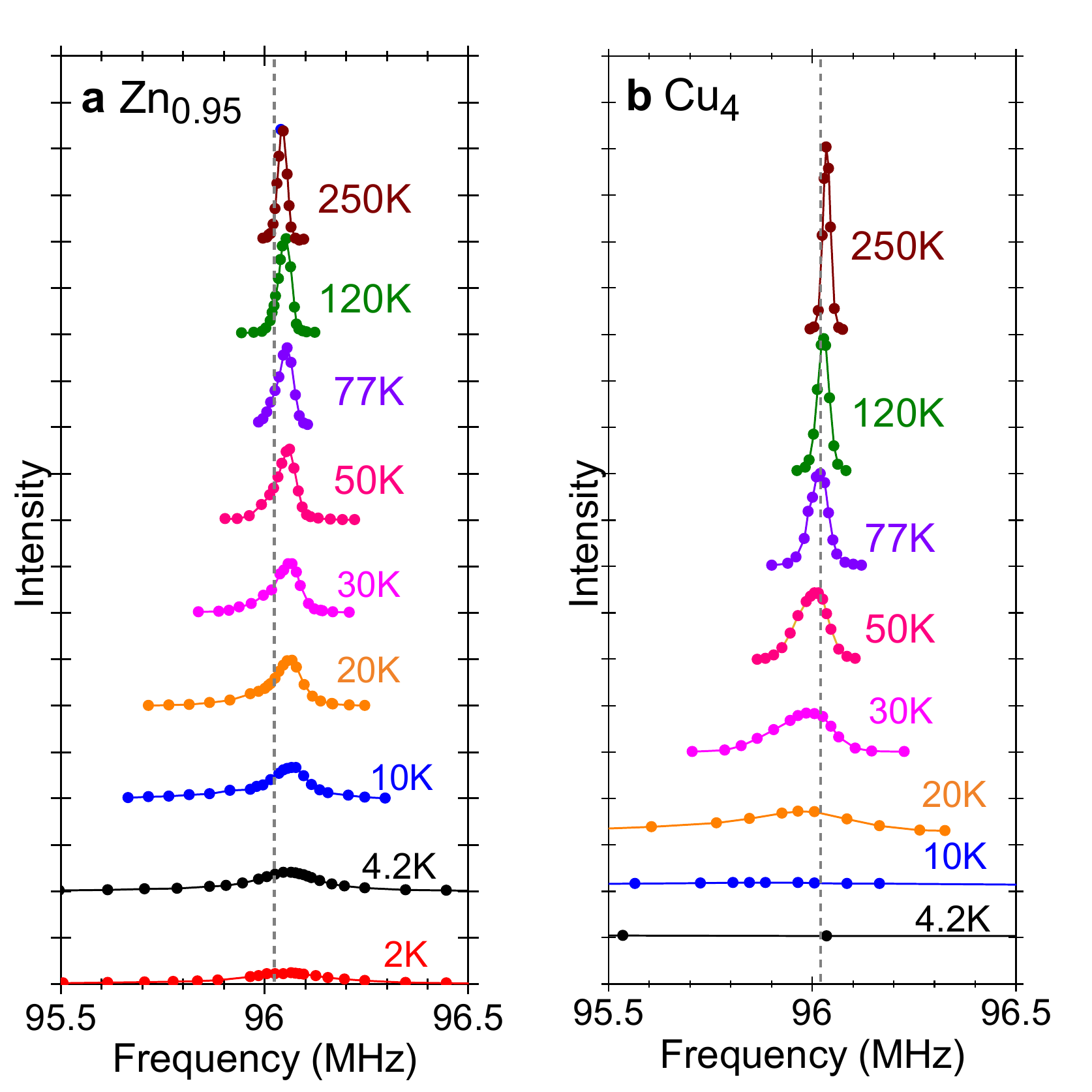}
		\caption{{\bf $^{19}$F NMR lineshapes}: Representative $^{19}$F NMR lineshapes of {\bf a}``Zn$_{0.95}$" (Zn$_{0.95}$Cu$_{0.05}$)Cu$_{3}$(OD)$_{6}$FBr and {\bf b} ``Cu$_{4}$" Cu$_{4}$(OD)$_{6}$FBr, both measured in $B_\text{ext} = 2.4$~T and normalized for Boltzmann factor.  For clarity, the vertical origin is shifted at different temperatures.  The lineshapes of Cu$_{4}$ below $T_{N}=15$~K is extremely broad due to static hyperfine magnetic field from ordered Cu moments.  The gray vertical dashed line represents the bare resonance frequency $^{19}f_\text{0}$ with no Knight shifts.}
		\label{fig:Fig2}
	\end{center}
\end{figure}

\begin{figure}
	\begin{center}
		\includegraphics[width=0.85\columnwidth]{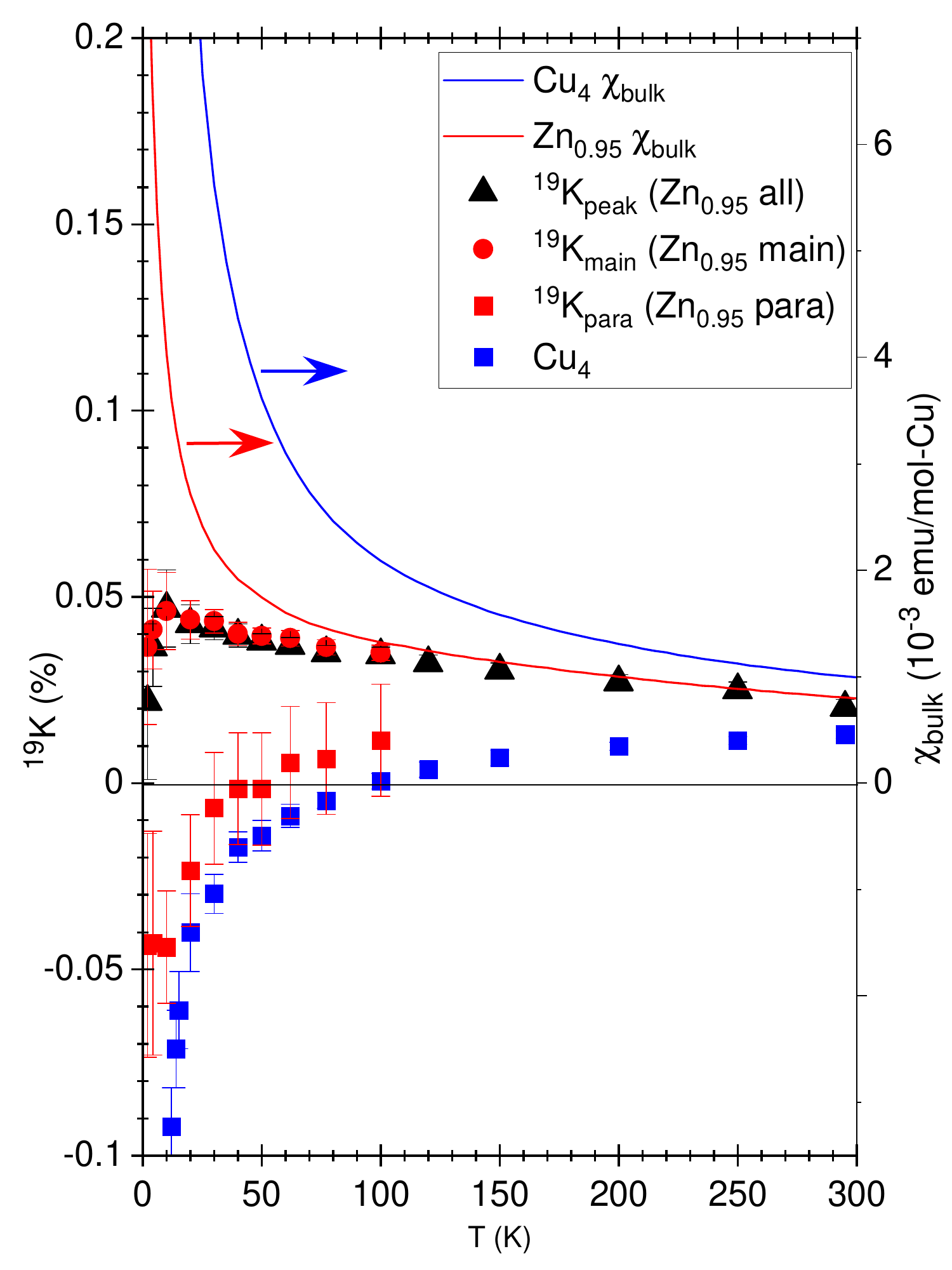}
		\caption{{\bf $^{19}$F NMR Knight shifts $^{19}K$}:  Summary of the Knight shift measured in $B_\text{ext}=2.4$~T for Zn$_{0.95}$: $^{19}K_\text{peak}$ as determined at the nominal peak frequency $^{19}f_\text{peak}$ of the overall lineshape ($\blacktriangle$), $^{19}K_\text{main}$ ($\color{red}{\bullet}$) and $^{19}K_\text{para}$ ($\color{red}{\blacksquare}$)  for the main and paramagnetic side peaks as determined from the deconvolution of $C(f;1/T_{1})$ based on double Gaussian fit.  For comparison, we show $^{19}K$ measured for the single peak observed for aniferromagnetic Cu$_{4}$ ($\color{blue}{\blacksquare}$).  The vertical error bars specify the maximum possible range of uncertainties.  Using the right axis, we also show the bulk magnetization data of $\chi_\text{bulk}$ measured in 2.4~T for Zn$_{0.95}$ (red) and Cu$_{4}$ (blue).}
		\label{fig:Fig3}
	\end{center}
\end{figure}

In Fig.~3, we summarize $^{19}K_\text{peak}$ ($\blacktriangle$) in comparison to $\chi_\text{bulk}$.  Our $^{19}K_\text{peak}$ results are similar to the earlier reports in \cite{Feng2017} and \cite{Wang2021}; $^{19}K_\text{peak}$ begins to decrease below $\sim20$~K toward zero.  An earlier report fitted the results with an activation form $^{19}K_\text{peak} \sim exp(-\Delta/k_{B}T)$ with a gap $\Delta$ \cite{Feng2017}, but we found that large experimental uncertainties of $^{19}K_\text{peak}$ caused by extremely broad lineshapes below 20~K make it rather difficult to rule out a linear temperature dependence expected for the gapless spin liquid picture based on Dirac Fermions \cite{Ran2007}.  We recall that $^{19}K_\text{peak}$ does not necessarily decrease to zero, unless all the 12 nn Cu sites of a given $^{19}$F site form spin singlets; but less than a half of Cu sites are actually involved in spin singlet formation in Zn$_{0.95}$ \cite{Wang2021}.  In fact, $^{19}K_\text{peak}$ exhibits a temperature dependence analogous to that of $\chi_\text{bulk}$ only down to $\sim 50$~K, where the latter begins to grow.   The broad $^{19}$F NMR linsehapes observed at low temperatures exhibit long tails stretching hundreds of kHz toward both higher and lower frequencies.  This means that a majority of $^{19}$F sites sense large, distributed hyperfine magnetic fields $B_\text{hyp}$ with both positive and negative signs, and only a small fraction of $^{19}$F sites exhibit diminishing $^{19}K_\text{peak}$ below 20~K.  

If there is no magnetic inhomogeneity, we expect $\chi^{i}_\text{local} = \chi_\text{bulk}$ for all sites, and hence the NMR lineshape would remain narrow.  In other words, the broadening of the lineshape indicates that $\chi^{i}_\text{local}$ develops a large distribution below $\sim50$~K.  Interestingly, this is the same temperature range, where the structural disorder freezes~\cite{Wang2022} as evidenced by the broadening of the $^{79}$Br NQR lineshapes; the oscillation on the transverse relaxation measurements also uncovered the signature of structural dimer formation below $\sim50$~K~\cite{Wang2022}.
 
We also measured $1/T_{1}$ at the peak frequency $^{19}f_\text{peak}$.  The magnitude of $1/T_{1}$ develops a significant distribution below $\sim160$~K where the lattice freezing sets in \cite{Wang2021}.  Accordingly, we resorted to deduce the stretched-fit value of $1/T_\text{1,str}$ based on the conventional stretched fit  of the recovery curve $M(t)$ as follows:
\begin{equation}
M(t) = M_{0} - A e^{-(t/T_\text{1,str})^{\beta}},
\label{eq:Stretch}
\end{equation}
where $M_\text{0}$ is the saturated nuclear magnetization, $A$ is the change in the nuclear magnetization immediately after the inversion $\pi$ pulse or saturation comb pulses, and $\beta$ is the phenomenological stretched exponent.  See Fig.~1g in \cite{Wang2021} for the similar results of $M(t)$ and the fit.  If magnetic inhomogeneity is negligibly small in Zn$_{0.95}$, we expect $\beta = 1$.  We emphasize that the stretched fit analysis of $M(t)$ is convenient but largely empirical, because it implicitly {\it assumes} a specific functional form for the density distribution function $P(1/T_{1})$ of $1/T_1$ \cite{Thayamballi1980,Lindsey1980,Itoh1986,JohnstonPRL2005,JohnstonPRB2006}.  Strictly speaking, it is justifiable only if $P(1/T_{1})$ may be represented with the modified Bessel function of the second kind \cite{JohnstonPRB2006}.  Empirically, $1/T_\text{1,str}$ is often a reasonable approximation of the center of gravity $1/T_\text{1,cg}$ of the true distributed values of $1/T_{1}$ as determined from ILT \cite{Singer2020,Arsenault2020,Takahashi2019,Wang2021,Mitrovic2008}, including the present case.

 \begin{figure}
	\begin{center}
		\includegraphics[width=1\columnwidth]{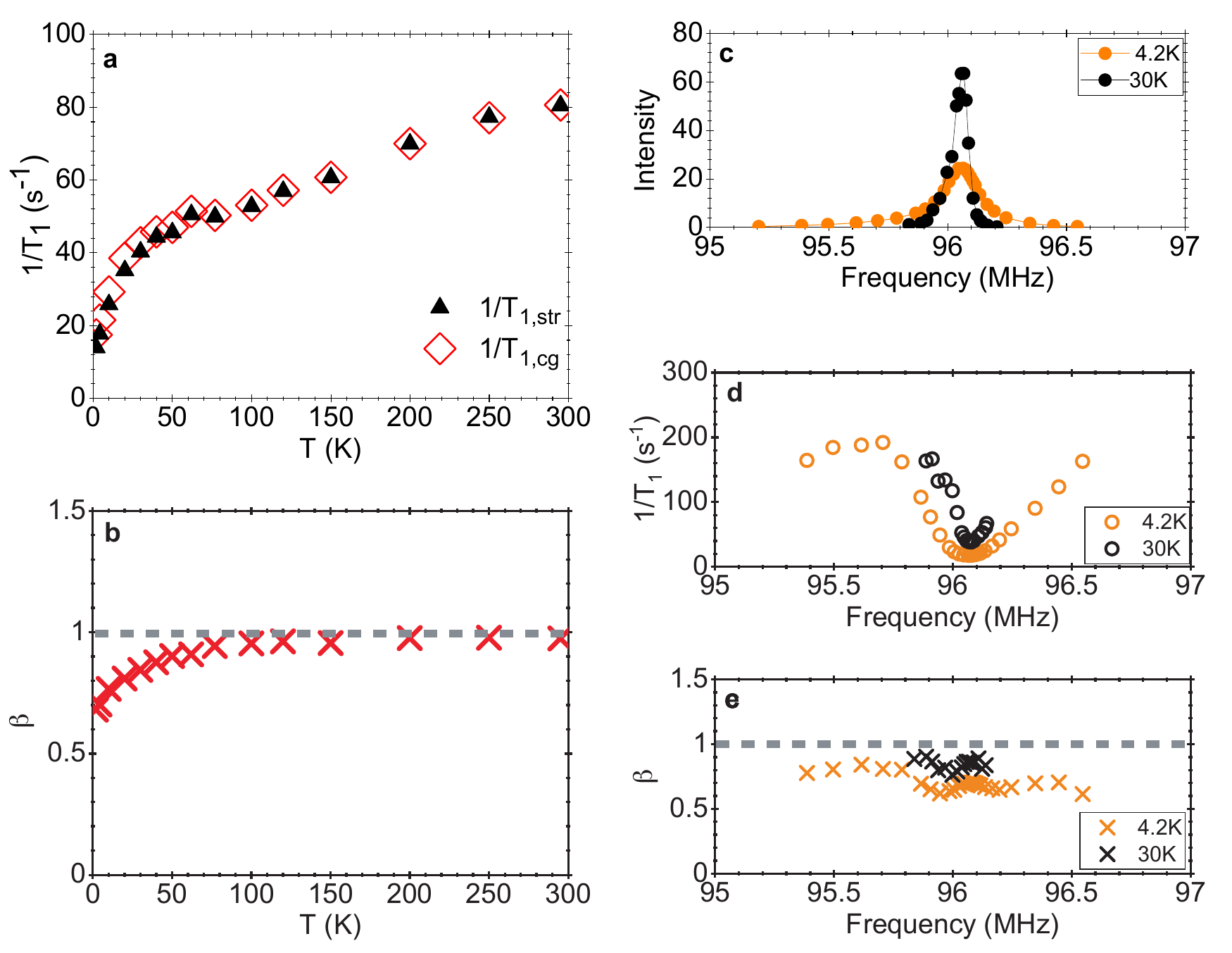}
		\caption{{\bf Conventional stretched fit results of $1/T_\text{1,str}$ in Zn$_{0.95}$}: {\bf a} Temperature dependence of $1/T_\text{1,str}$ observed at $^{19}f_\text{peak}$ for Zn$_{0.95}$.  For comparison, we also show the center of gravity $1/T_\text{1,cg}$ of the distributed $1/T_1$ as determined from ILT.  {\bf b} Temperature dependence of the stretched exponent $\beta$ observed at the peak.  {\bf c} Point-by-point $^{19}$F NMR lineshape at 30~K and 4.2~K.  {\bf d-e} $1/T_\text{1,str}$ and $\beta$ measured at each data point in {\bf c}.  Notice that $1/T_\text{1,str}$ is enhanced at both higher and lower frequencies.   $\beta$ deviates from 1 more significantly near the center of the lineshapes, because the fast and slow contributions are superimposed.
}
		\label{fig:Fig4}
	\end{center}
\end{figure}

We summarize the temperature dependences of $1/T_\text{1,str}$ and $\beta$ observed at $^{19}f_\text{peak}$ in Fig.4a and b.  These results are very similar to our earlier work conducted in a lower magnetic field 0.72~T \cite{Wang2021}.  $1/T_\text{1,str}$ smoothly decreases with temperature, and does not exhibit an activation behavior expected for fully and homogeneously gapped QSL.  On the other hand, the power law behavior expected for the gapless QSL formed by Dirac Fermions \cite{Ran2007} is not observed, either.  A key consideration in interpreting the $1/T_{1}$ results is that, again, $^{19}$F sites are subjected to the fluctuations of the transferred hyperfine fields $B_\text{hyp}$ from 12 nn Cu$^{2+}$ sites.  Since the absolute upper bound of the fraction of spin singlets occupying the Cu sites is $F_\text{singlet}\sim50$\% in Zn$_{0.95}$ \cite{Wang2021}, many of the $^{19}$F sites are under the influence of paramagnetic Cu spins that are not involved in spin singlets.  This explains why the results for $1/T_\text{1,str}$ as well as $\chi_\text{local}$ do not meet the naive expectations for purely gapped or purely gapless QSL.

We also measured $1/T_\text{1,str}$ at various  frequencies within the extremely broad  lineshape.  We summarize the representative results of the frequency dependence of $1/T_\text{1,str}$ and $\beta$ in Fig.4c-e.  $1/T_\text{1,str}$ is larger at both lower and higher frequency sides of the peak.   This suggests that the $1/T_{1}$ relaxation process at $^{19}$F sites in the tailed sections of the broad lineshape is enhanced by low energy spin excitations associated with the source of the hyperfine magnetic fields that gives rise to the line broadening.  But we are unable to pinpoint their origin based on these measurements.\\  

\noindent{\bf Two-dimensional NMR data $C(f;1/T_{1})$}\\

\indent Proximate quantum spin liquid materials that do not undergo magnetic long range order are often structurally disordered and exhibit significantly distributed $1/T_1$.  In our recent work, we overcame the difficulties in conventional NMR data acquisition protocols outlined in the previous section by using the ILTT$_1$ analysis technique \cite{Singer2020,Arsenault2020}.  In the ILT analysis of the  distributed $1/T_1$, one only assumes that $1/T_1$ has a spatial distribution, and each nuclear spin relaxes exponentially with the j-th value $1/T_{1j}$ with the probability density $P(1/T_{1j})$.  Utilizing Tikhonov regularization, one can optimize the fit of the experimentally observed recovery curve $M(t)$ with a summation of single exponentials with distributed $1/T_1$,
\begin{equation}
M(t) = \sum_{j} M_\text{0} \left  [1 - 2 e^{-t/T_{1j}}\right] P(1/T_{1j}),
\label{eq:ILT}
\end{equation}
where  the summation of the probability $P(1/T_{1j})$ is normalized to 1.  See \cite{Mitchell2012,Singer2020,Arsenault2020} for the details of the numerical  ILT procedures.

In Fig.5a, we summarize representative $P(1/T_{1})$ results deduced for the distribution of $1/T_1$ in Zn$_{0.95}$ measured at $^{19}f_\text{peak}$ in $B_\text{ext}=2.4$~T.  These $P(1/T_{1})$ results are similar to our previous findings in $B_\text{ext}=0.72$~T \cite{Wang2021}, and indicate that $1/T_1$ has a roughly symmetrical distribution above $\sim 100$~K.  The center of gravity $1/T_\text{1,cg}$ of the distribution of $P(1/T_{1})$ summarized in Fig.4a is similar to $1/T_\text{1,str}$, as noted above.  Below $\sim100$~K, $P(1/T_{1})$ begins to develop a tail toward larger values of $1/T_1$.  This means that some of the $^{19}$F nuclear spins resonating around $^{19}f_\text{peak}$ have enhanced $1/T_1$, but the majority of $^{19}$F nuclear spins around $^{19}f_\text{peak}$ still relax with the slower values comparable to $1/T_\text{1,str}$.  We emphasize that the conventional stretched fit analysis alone cannot clarify such qualitative changes in the distribution of $1/T_1$.    

 \begin{figure}
	\begin{center}
		\includegraphics[width=1\columnwidth]{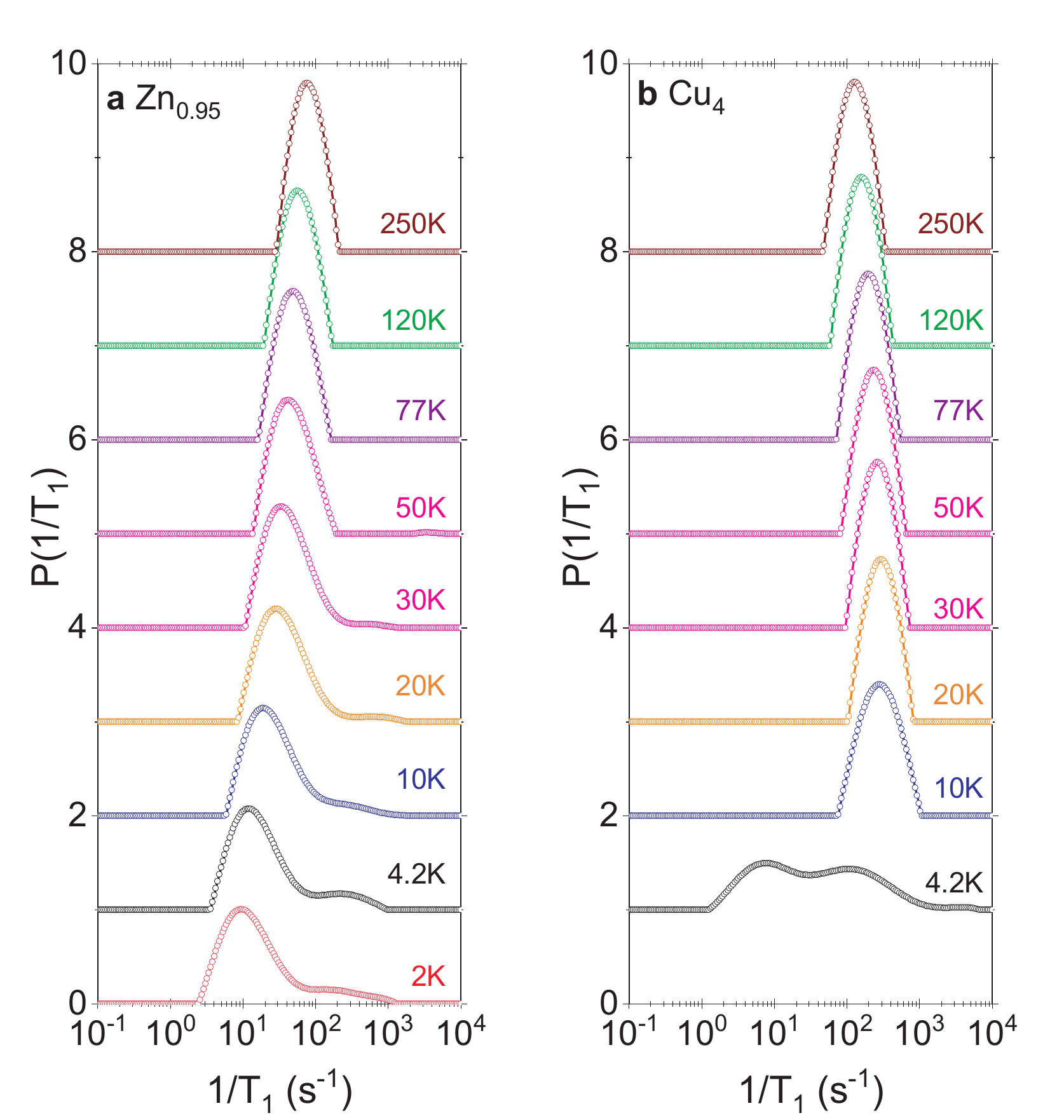}
		\caption{{\bf Density distribution function $P(1/T_{1})$}: {\bf a} Representative results of $P(1/T_{1})$ deduced at the peak frequency $^{19}f_\text{peak}$ of the $^{19}$F NMR linehapes observed for Zn$_{0.95}$.  The integral of $P(1/T_{1})$ is normalized to 1 (one).   {\bf b} Corresponding results observed for Cu$_{4}$.  Notice that the side peak of $P(1/T_{1})$ observed for Zn$_{0.95}$ with $1/T_{1}\sim 300$~s$^{-1}$  is absent for Cu$_{4}$.  The double peak structure observed at 4.2~K is caused by antiferromagnetic long range order.  We fixed Tikhonov regularization parameter for ILT as $\alpha=1$ to achieve the same effective resolution.  The distributions of $1/T_1$ is negligibly small with $\beta \simeq 1$ above $\sim160$~K for Zn$_{0.95}$ and above $T_{N}=15$~K  for Cu$_{4}$, hence the apparent width of $P(1/T_{1})$ in these regions is set primarily by the effective resolution of our ILT with fixed $\alpha=1$.  (See Supplemental Materials of \cite{Singer2020,Wang2021} for details about the effective resolution of the ILT analysis.)
}
		\label{fig:Fig5}
	\end{center}
\end{figure}

Successful demonstration of the distribution $P(1/T_{1})$ of $1/T_1$ in Zn$_{0.95}$ \cite{Wang2021} was a major step toward clarifying the nature of disorder  in  kagome materials.  However, we were unable to identify the origin of the $^{19}$F sites with enhanced $1/T_1$, nor can we estimate their population based on such a one-dimensional NMR data set alone even with ILT.  Notice that we were measuring two sets of one-dimensional distributions separately in these measurements: (a) the distribution of the resonant frequency $^{19}f$ and hence $^{19}K$ in the form of the NMR lineshape, which provides us with the histogram of the distribution of $\chi_\text{local}$ averaged over 12 Cu$^{2+}$ sites nn to each $^{19}$F site; (b) the distribution of $1/T_1$ in the form of $P(1/T_{1})$ at the peak frequency $^{19}f_\text{peak}$.  What is still missing is the correlation between the distributions in static local spin susceptibility $\chi_\text{local}$ and local spin dynamics reflected on $1/T_1$. 

Here, we take one more step forward to clarify the nature of disorder by advancing the ILTT$_1$ analysis technique.  In addition to $P(1/T_{1})$ at the peak frequency $^{19}f_\text{peak}$ of the lineshape, we also measure $M(t)$ and use ILT to deduce $P(1/T_{1})$ at other frequencies within the broadened lineshape, and thereby {\it correlate the histograms of $^{19}K$ and $1/T_1$ distributions}.  In Fig.~6, we illustrate how this two-dimensional NMR data acquisition scheme works.  The top panel Fig.~6a shows the conventional overall NMR lineshape at 30~K.  The broad lineshape peaked at $^{19}f_\text{peak} \simeq 96.06$~MHz represents the histogram of the total hyperfine magnetic field at $^{19}$F sites transferred from the 12 nn Cu sites.  The result of $P(1/T_{1})$ measured at $^{19}f_\text{peak}$, which we now represent as $C(^{19}f_\text{peak};1/T_{1})$ to show the specific frequency, is shown in the bottom panel Fig.~6c underneath the peak of the total lineshape in Fig.~6a.  Note that this is the same $P(1/T_{1})$ curve at 30~K presented in Fig.~5a, rotated clockwise by 90$^{\circ}$.  The integrated area of the $C(^{19}f_\text{peak};1/T_{1})$ as a function of $1/T_1$ is normalized to 1, as before.  

 \begin{figure}
	\begin{center}
		\includegraphics[width=1\columnwidth]{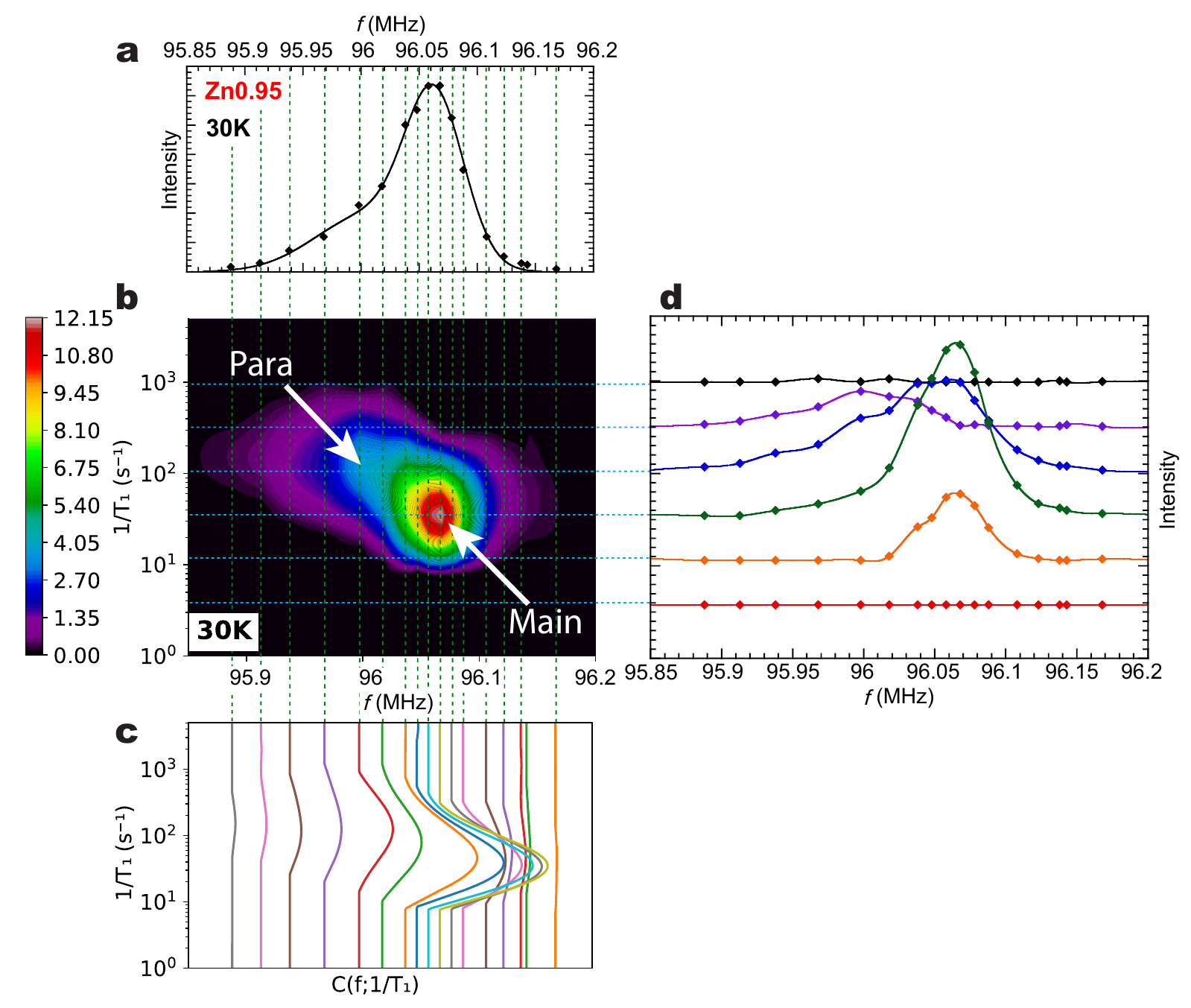}
		\caption{{\bf Two dimensional correlation map $C(f;1/T_{1})$ of Zn$_{0.95}$ and $1/T_1$-resolved NMR lineshapes}: {\bf a} Conventional one-dimensional NMR lineshape of all $^{19}$F nuclear spins observed for Zn$_{0.95}$ at 30~K. {\bf b} Example of the color-contour map of the frequency $f$ vs. $1/T_1$ correlation function $C(f;1/T_{1})$ at 30~K deduced from the ILT measurements of $P(1/T_{1})$ at various frequencies within the lineshape.   A side peak marked as Para emerges at low temperatures centered around $f=96.02$~MHz or below, with enhanced $1/T_{1}\simeq 100$~s$^{-1}$. {\bf c}  Vertical slice of the color contour map of $C(f;1/T_{1})$ at a fixed frequency $f$ is nothing but the one-dimensional $P(1/T_{1})$ result at that frequency.  For example, $C(f;1/T_{1})$ sliced vertically at $^{19}f_\text{peak} = 96.06$~MHz is $P(1/T_{1})$ at 30~K presented in Fig.5a.  {\bf d} Horizontal slice of $C(f;1/T_{1})$ at a fixed value of $1/T_1$ yields the $1/T_{1}$-resolved NMR lineshape with the contributions from only the $^{19}$F nuclear spins with that particular value of  $1/T_1$.  Note that the $^{19}$F NMR $1/T_{1}$-resolved lineshape for $1/T_{1} = 300$~s$^{-1}$ is peaked at 96.02~MHz, which is lower than the main peak frequency $^{19}f_\text{peak} \simeq 96.06$~MHz.  The Knight shift $^{19}K_\text{para}$ arises from the former. 
}
		\label{fig:Fig6}
	\end{center}
\end{figure}

Next, we repeat the measurement of the recovery curves $M(t)$ at different frequencies $f$ within the broad lineshape, and deduce the corresponding $C(f;1/T_{1})$ between $f =95.9$~MHz and 96.15~MHz, as also shown in Fig.~6c.  The integrated area of $C(f;1/T_{1})$ at each frequency is normalized to the relative intensity of the one-dimensional NMR lineshape in Fig.6a.  For example, the overall $^{19}$F signal intensity at 96.02~MHz in Fig.~6a is only $\sim50$\% of the peak intensity at $^{19}f_\text{peak}\simeq96.06$~MHz, hence the integrated area of $C(\text{96.02MHz};1/T_{1})$ is normalized to $\sim0.5$.  Once we deduce  multiple $C(f;1/T_{1})$ curves across the broad NMR lineshape in panel Fig.6c, we can generate the two-dimensional color contour plot of $C(f;1/T_{1})$ as a function of {\it both} $f$ {\it and} $1/T_1$ and normalize its integral in the two-dimensional parameter space to 1.  We present the normalized color contour plot of $C(f;1/T_{1})$ in panel Fig.6b.   $C(f;1/T_{1})$ represents the two-dimensional correlation map between the distributed resonance frequency $f$ {\it and} the distributed low energy spin excitations reflected on $1/T_1$.  Therefore, the magnitude of $C(f;1/T_{1})$ at specific values of $f$ {\it and} $1/T_1$ represents the relative probability of a $^{19}$F nuclear spin to have particular values of $f$ {\it and} $1/T_1$.  

These two-dimensional NMR data of $C(f;1/T_{1})$ require an order of magnitude longer data acquisition time in comparison to the conventional one-dimensional NMR data, because we need to repeat $1/T_1$ measurements at various frequencies with high precision required by ILT.  But the results of the correlation map $C(f;1/T_{1})$ provide much richer information.  For example, we can slice the color contour map in Fig.~6b horizontally at a given value of $1/T_{1}'$, and plot $C(f; 1/T_{1}')$ as a function of $f$, as summarized in the right panel Fig.~6d.  Then the resulting curves represent the NMR lineshapes for nuclear spins with the specific values of $1/T_{1}'$.  Unlike the conventional NMR lineshape in panel Fig.6a measured from the sum of {\it all} $^{19}$F nuclear spins with various $1/T_1$ values, the NMR lineshapes in Fig.~6d are resolved for different $1/T_1$ values.  We opt to coin them as {\it ILTT$_1$ resolved NMR lineshapes}.  Notice that the ILTT$_1$ resolved NMR lineshape with $1/T_{1}\simeq300$~s$^{-1}$ is centered around 96.02~MHz (purple curve in Fig.~6d), whereas the majority of the $^{19}$F sites are concentrated around the main peak near 96.05~MHz with much slower values of $1/T_{1}\simeq 30$~s$^{-1}$ (green curve in Fig.~6d).  To the best of our knowledge, this is the first time to successfully resolve the inhomogeneously broadened NMR lineshape for different values of  $1/T_{1}$, although narrow NMR lineshapes with homogeneous broadening were previously resolved with the aid of fast Fourier transform \cite{Sun2005}.\\

\noindent{\bf Temperature dependence of $C(f;1/T_{1})$}\\

In Fig.~7a-d, we summarize the representative results of $C(f;1/T_{1})$ observed for Zn$_{0.95}$ at various temperatures.  See Supplemental Materials for the complete set of $C(f;1/T_{1})$ data.  Supplemental files also include a motion picture (movie) summary of the evolution of $C(f;1/T_{1})$ as a function of temperature.  Notice that $C(f;1/T_{1})$ has a nearly symmetrical oval shape around 250~K, because both $f$ and $1/T_1$ have only a minor distribution centered around their central value.   This main peak remains fairly narrow along both the $f$ and $1/T_1$ axes at least down to 10~K.  But a distinct component emerges below $\sim120$~K on the upper left hand side of the main peak, and gradually splits off.  In what follows, we call the emergent side peak with larger values of $1/T_1$ as the paramagnetic peak for the reasons to become clear below.  The two dimensional distribution of the paramagnetic peak continuously grows along the $f$ axis toward 2~K, but the growth of the distribution along the $1/T_1$ axis appears less significant at low temperatures.  

 \begin{figure}
	\begin{center}
		\includegraphics[width=1\columnwidth]{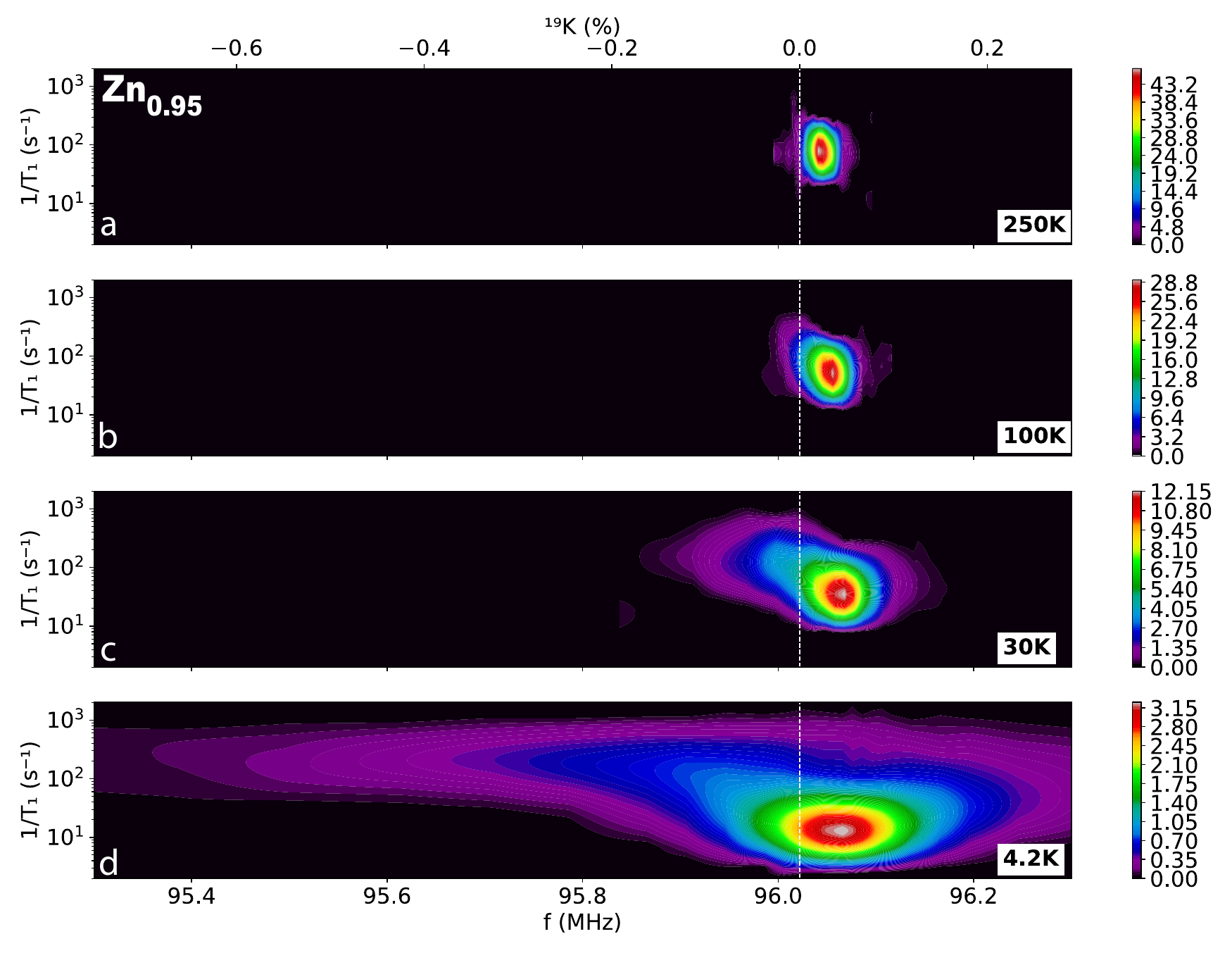}
		\includegraphics[width=1\columnwidth]{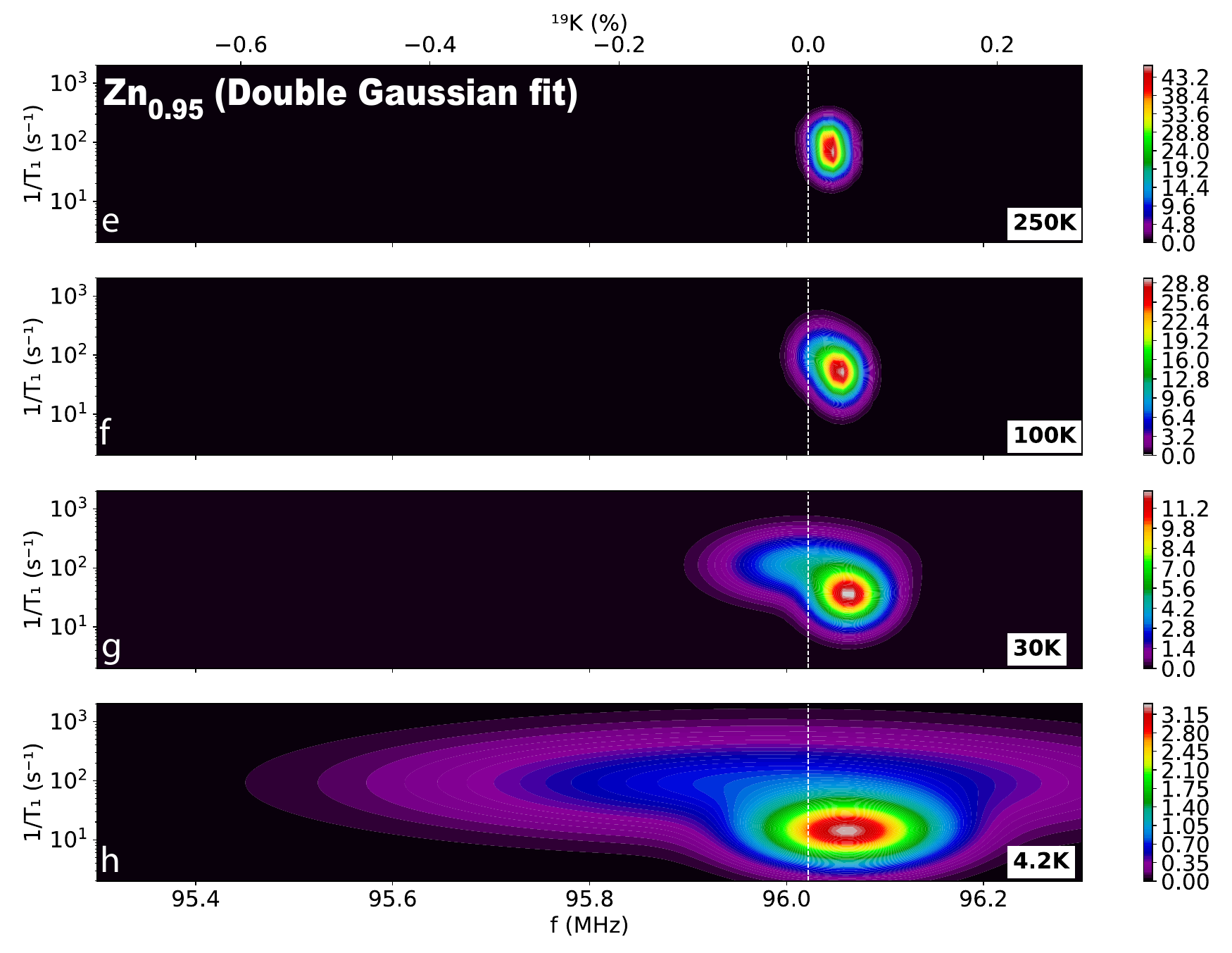}
		\caption{{\bf Evolution of $C(f;1/T_{1})$ for Zn$_{0.95}$}: {\bf a-d} Representative $C(f;1/T_{1})$ results at various temperatures observed for Zn$_{0.95}$.  Vertical dashed line marks the zero Knight shift frequency $^{19}f_\text{0}$.  See Supplemental Information for the complete set of the data as well as the animation (movie) of the temperature evolution of $C(f;1/T_{1})$.  {\bf e-h} $C(f;1/T_{1})$ at corresponding temperatures, deduced from the summation of the two Gaussian functions in the two-dimensional $f$ and $1/T_1$ space that best fit the experimental results in {\bf a-d}.  Note that above 100~K, the two-dimensional gaussian does not capture the behavior of the side peak as it is merged into the main peak.
}
		\label{fig:Fig7}
	\end{center}
\end{figure}

In order to separate the contributions of the main and paramagnetic peaks, we deconvoluted $C(f;1/T_{1})$ into two separate two-dimensional Gaussian functions of $f$ and $1/T_{1}$ by conducting the least $\chi^{2}$ fit in the two-dimensional parameter space.  See Supplementary Materials for the details of the fit.  We show the color contour map of the summation of these two Gaussian components in Fig.~7e-h.  The two component fit reproduces the experimental data in Fig~7a-d fairly well.  The paramagnetic side peak merges into the main peak at higher temperatures, and we were unable to clearly resolve them above 100~K.  In Fig.~1d, we summarize the temperature dependence of the fraction $F_\text{para}$ of the $^{19}$F sites involved in the paramagnetic side peak estimated from the integral of the two-dimensional Gaussian peak in the $f$-$1/T_1$ space.   $F_\text{para}$ gradually grows with decreasing temperature, and reaches as large as $F_\text{para} \sim 0.6$ at 2~K.  We emphasize that the faint signature of the paramagnetic side peak seems to persist even at 150~K or above, as shown in Supplemental Figure 3.  It is consistent with our earlier finding that the $^{2}$D and $^{17}$O sites nn to the interlayer defects are distinct even at 295~K in herbertsmithite \cite{Imai2011,Fu2015}.  On the other hand, the estimation of $F_\text{para}$ was  difficult above 100~K due to the unstable nature of the two dimensional fit of the small side peak involving a large number of free parameters.

From the central values of the two-dimensional Gaussian function for the paramagnetic peak, we estimated the central values of the distributed $^{19}K_\text{para}$ and $1/T_{1}^\text{para}$, whereas $^{19}K_\text{main}$ and $1/T_{1}^\text{main}$ were estimated from the two-dimensional Gaussian function of the main peak; we summarize these results in Fig.~3 and Fig.~8.  The paramagnetic peak is somewhat asymmetrical along the frequency axis with slightly more weight along the lower frequency side, but this does not significantly affect our estimation of the central value of $^{19}K_\text{para}$.  The central values of $^{19}K_\text{main}$ and $1/T_{1}^\text{main}$ for the main peak are close to the nominal peak values of the one dimensional data discussed in the previous section.   \\

 \begin{figure}
	\begin{center}
		\includegraphics[width=0.95\columnwidth]{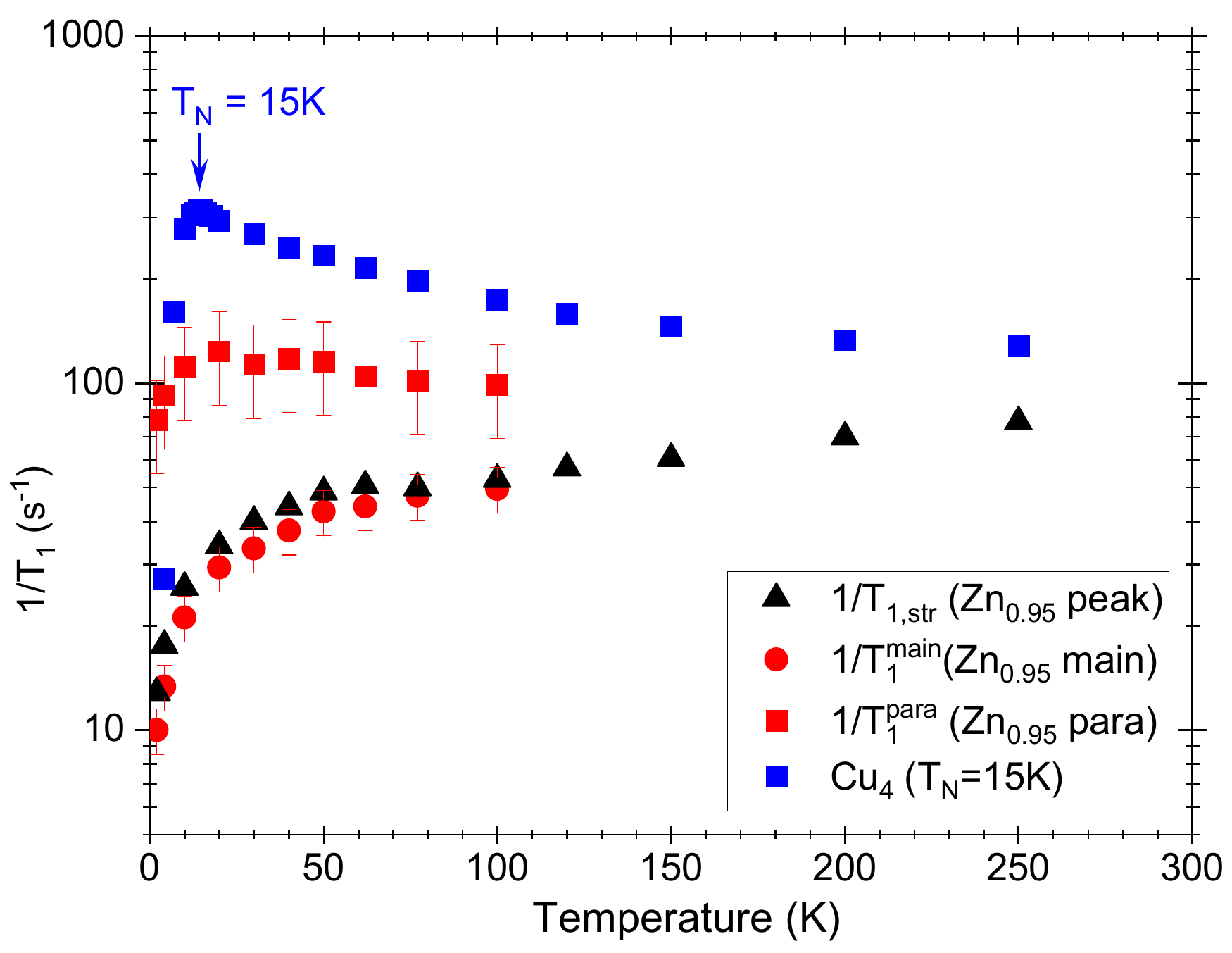}
		\caption{{\bf Temperature dependences of $1/T_{1}$}: The central value $1/T_\text{1}^\text{main}$ (\textcolor{red}{$\bullet$}) of the main peak and $1/T_\text{1}^\text{para}$  (\textcolor{red}{$\blacksquare$}) of the paramagnetic peak observed for Zn$_{0.95}$, as estimated from the peak of the double Gaussian fit of $C(f;1/T_{1})$.  The vertical error bars specify the maximum possible range of uncertainties.   For comparison, we also show $1/T_\text{1,str}$ observed at the nominal peak frequency $^{19}f_\text{peak}$ of Zn$_{0.95}$ ($\blacktriangle$, same stretched fit results as in Fig.4a) and Cu$_4$ (\textcolor{blue}{$\blacksquare$}).
		}
		\label{fig:Fig8}
	\end{center}
\end{figure}

\noindent{\bf DISCUSSIONS}\\

\indent Our two-dimensional correlation maps $C(f;1/T_{1})$ and ILTT$_1$ resolved $^{19}$F NMR lineshapes revealed that two distinct types of $^{19}$F sites exist in Zn$_{0.95}$ at low temperatures: (i) $^{19}$F sites with characteristic signatures of paramagnetic spins, {\it i.e.} enhanced Knight shift $^{19}K_\text{para}$ due to growing spin polarization, and large temperature independent $1/T_{1}^\text{para}$ induced by gapless low energy spin excitations obeying $\chi'' \sim 1/T$.  The volume fraction $F_\text{para}$ of these $^{19}$F sites reaches as much as $\sim60$~\% at 2~K, indicating that spin polarized domains gradually extends in space, as schematically shown in Fig.~1c.  (ii) The main intrinsic $^{19}$F sites that are mostly immune from these effects, suggesting that they are spatially more distanced from the source of the spin polarization and more reflective of the intrinsic behavior of the kagome planes.  These two different types of $^{19}$F sites become indistinguishable above $\sim120$~K within our experimental resolutions.

Interestingly, the negative Knight shift $^{19}K_\text{para}$ of the paramagnetic peak shows qualitatively the same behavior as the negative Knight shift observed at the nn $^{2}$D and $^{17}$O sites of the 15\% interlayer Cu$^{2+}$ defect spins occupying the Zn$^{2+}$ sites in herbertsmithite \cite{Imai2011,Fu2015}.  This strongly suggests that the paramagnetic $^{19}$F sites are also located in the vicinity of the 5\% interlayer Cu$^{2+}$ defects of Zn$_{0.95}$.  In order to test this scenario, we repeated two-dimensional $^{19}$F NMR measurements in a deuterated powder sample of barlowite Cu$_{4}$(OD)$_{6}$FBr.  Cu$_4$ is the parent antiferromagnetic phase of Zn$_{0.95}$ with Cu$^{2+}$ spins occupying all the interlayer Zn$^{2+}$ sites in Zn$_{0.95}$, and undergoes a N\'eel transition into antiferromagnetically ordered state at $T_{N} = 15$~K \cite{Ranjith2018,Tustain2018,Smaha2020}.

We summarize the $^{19}$F NMR results of Cu$_4$ in Fig.2b, 3, 5b, 8, and 9 in comparison to the results for Zn$_{0.95}$.  Our one-dimensional NMR data for Cu$_4$ are similar to earlier $^{19}$F NMR measurements conducted in a limited temperature range below 100~K for a protonated powder sample of Cu$_4$ \cite{Ranjith2018}.    Notice that two-dimensional correlation maps $C(f;1/T_{1})$ of Cu$_4$  in Fig.~9 show no signature of a split-off peak.  Instead, Cu$_4$ exhibits only one type of $^{19}$F NMR peak in the entire paramagnetic state above $T_{N}$; this is consistent with the crystal structure of Cu$_4$ with a crystallographically unique $^{19}$F site.  

 \begin{figure}
	\begin{center}
		\includegraphics[width=1\columnwidth]{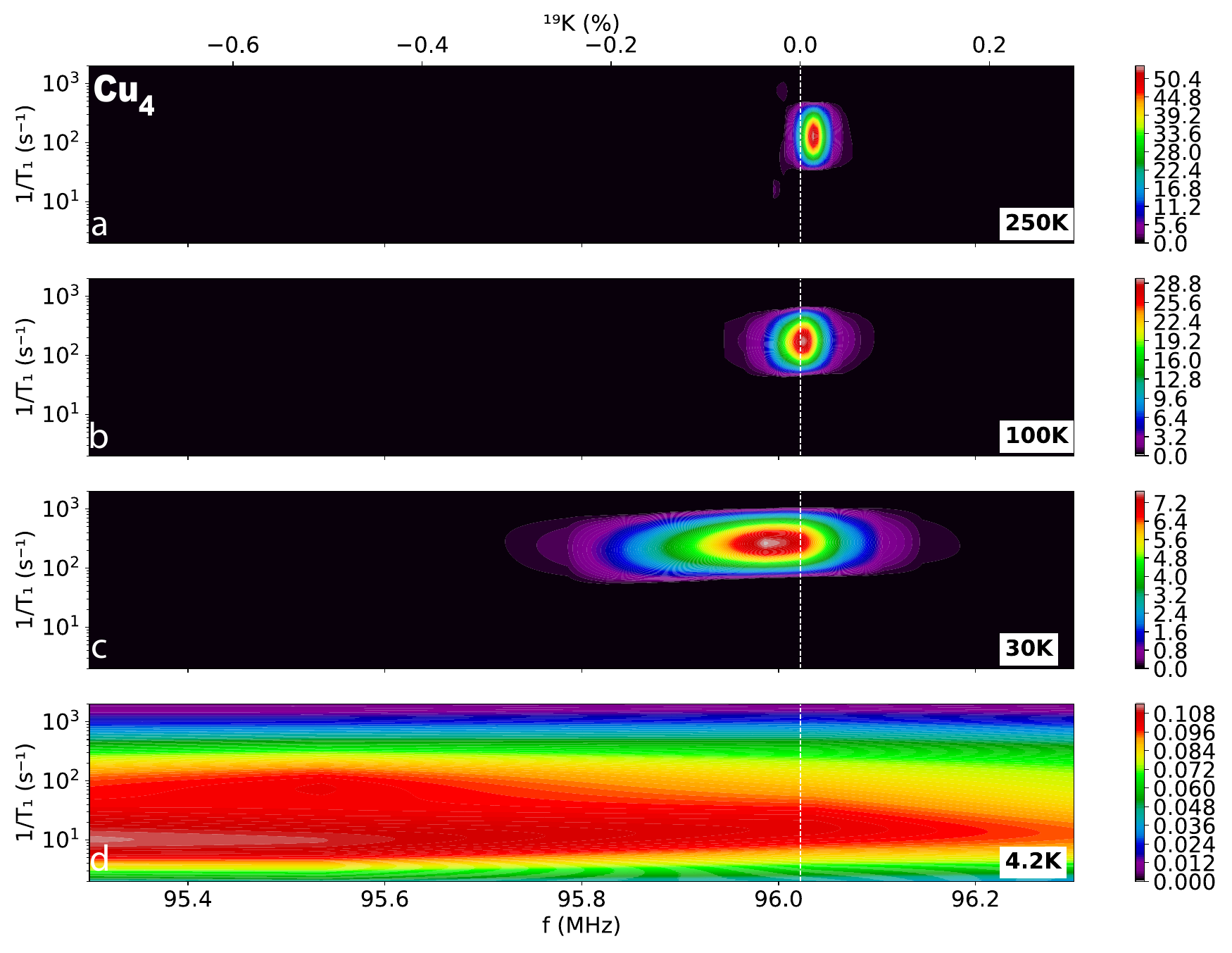}
		\caption{{\bf Two dimensional correlation map $C(f;1/T_{1})$ of Cu$_{4}$}: {\bf a-d} The two-dimensional correlation map $C(f;1/T_{1})$ observed for Cu$_{4}$ at representative temperatures.  The $^{19}$F NMR lineshape gradually broadens below 250~K as shown in Fig.~2b toward $T_{N}=15$~K, but the distribution of $1/T_{1}$ is always minimal and roughly symmetrical at all frequencies in the paramagnetic state.  Unlike Zn$_{0.95}$, there is no hint of a distinct split off peak above $T_N$.  The drastic change of $C(f;1/T_{1})$ at 4.2~K is due to antiferromagnetic long range order.}
		\label{fig:Fig9}
	\end{center}
\end{figure}

 \begin{figure}
	\begin{center}
		\includegraphics[width=0.9\columnwidth]{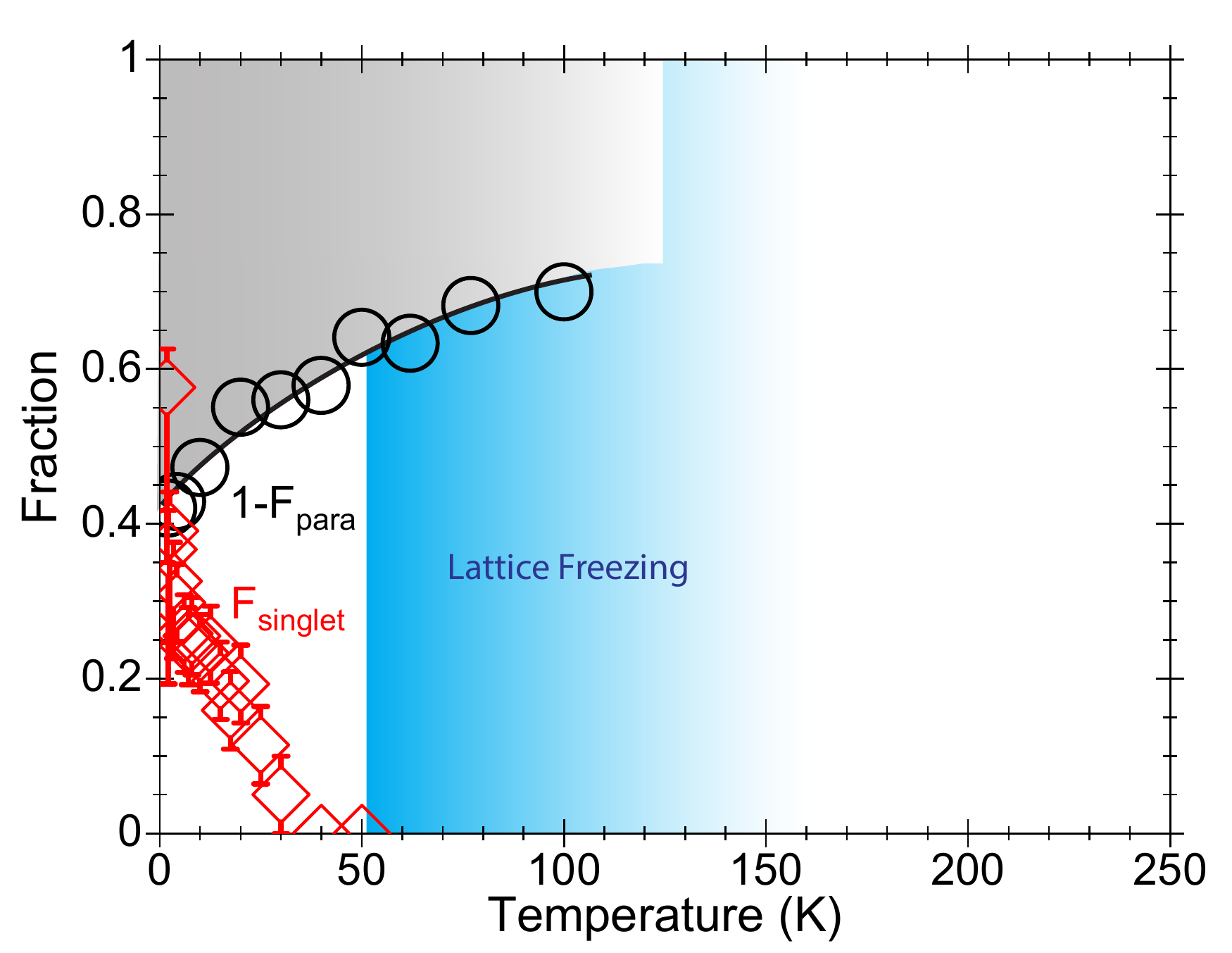}
		\caption{{\bf Volume fraction of paramagnetic and spin singlet sites}:  $1-F_\text{para}$ ($\circ$, obtained from the results in Fig.~1d) represents the shrinking volume fraction of Zn$_{0.95}$, which is {\it not} within the spin polarized domain and hence could exhibit intrinsic kagome spin liquid behavior without the perturbation caused by interlayer Cu$^{2+}$ defects occupying the Zn$^{2+}$ sites.  Notice that only $\sim40$\% of such a volume remain at 2~K in Zn$_{0.95}$.  Also shown (\textcolor{red}{$\lozenge$}) is the {\it upper bound} of the spin singlet fraction $F_\text{singlet}$ of Cu sites, estimated from the observable $^{63}$Cu NQR signals at the fixed pulse separation time $\tau = 6$~$\mu$s \cite{Wang2021} (The lower error bars below 6 K represent the lower bound
of $F_\text{singlet}$, estimated based on the assumption that the $^{63}$Cu signal loss affects only the paramagnetic component. All other error bars correspond to the absolute maxima or minima of $F_\text{singlet}$ \cite{Wang2021}.)  Blue shading schematically shows the temperature range where the freezing of the structural distortion sets in ($\sim 160$~K) and completes ($\sim 50$~K).
}
		\label{fig:Fig10}
	\end{center}
\end{figure}

A striking finding here is that $^{19}K_\text{para}$ observed for Zn$_{0.95}$ shows nearly identical behavior as the unique $^{19}$F sites in antiferromagnetic Cu$_4$ above $T_N$, as shown in Fig.~3.    This contrasts with the behavior of $^{19}K_\text{main}$ and reinforces our conclusion that the paramagnetic $^{19}$F sites observed for Zn$_{0.95}$ are in the close proximity with interlayer Cu$^{2+}$ defects occupying the Zn$^{2+}$ sites, where the structural and magnetic environment is locally similar to Cu$_4$'s.  Since $\chi_\text{bulk}$ of Cu$_4$ shows strong enhancement below $\sim100$~K, naturally one expects analogous enhancement of the local spin susceptibility $\chi_\text{local}$ in the vicinity of Cu$^{2+}$ interlayer defects of Zn$_{0.95}$.  The negative sign of the hyperfine magnetic field $B_\text{hyp}$ leads to strongly negative $^{19}K_\text{para}$ in Zn$_{0.95}$.  This further suggests that the conclusions from prior one dimensional $^{19}$F NMR studies that neglect the affects of impurities should be reassessed \cite{Feng2017}. 

While our $^{19}$F NMR measurements establish the gradual growth of spin polarized domains induced in the vicinity of the Cu$^{2+}$ interlayer defects, our measurements do not reveal the exact spin configuration in these domains. The growth of staggered spin polarization within the kagome planes was previously proposed to account for the broadening of $^{17}$O NMR lineshapes in herbertsmithite, but it was attributed to the Zn$^{2+}$ anti-site defects within the kagome planes \cite{Olariu2008,Khunita2020}.  Our findings here indicate that such spin vacancies within the kagome planes are not required to account for the NMR line broadening in these kagome materials; after all, Cu$_{4}$ has no Zn$^{2+}$ ions in its composition.  Moreover, the upper bound of the Zn$^{2+}$ anti-site defect population is as little as $\sim1$\% \cite{Freedman2010,Smaha2020}.  At high temperatures ($T > 160$~K) dominated by short range spin fluctuations, it is striking that the behavior of Zn$_{0.95}$ ($^{19}K_\text{main}$ and $1/T_1^{main}$ decrease upon cooling) clearly differs from Cu$_{4}$ ($^{19}K$ and $1/T_1$ increase upon cooling).  This hints at an approach to two different ground states.

Another important point to emphasize is that the fraction of the paramagnetic $^{19}$F sites in Fig.~1d gradually grows below $\sim100$~K to as large as $F_\text{para}\sim60$\% toward the base temperature of 2~K.  In Fig.1c, we use a circle with orange-to-yellow shading to schematically present the spin polarized domain surrounding a Cu$^{2+}$ defect spin occupying the interlayer Zn$^{2+}$ site in the middle.  Note that 12 $^{19}$F sites in the field of view belong to the domain out of 20 $^{19}$F sites, corresponding to 60\%.  Assuming that the inter-layer defect sites are randomly located without forming a cluster, Fig.1c suggests that kagome Cu$^{2+}$ sites within the range up to third or fourth nearest-neighbor may be affected by each inter-layer Cu$^{2+}$ defect spin at 2~K.  As $F_\text{para}$ decreases at higher temperatures, the range of the spin polarized domain smoothly decreases.

In our recent work, we demonstrated based on $^{79}$Br NQR measurements that the crystal structure develops local distortions below $\sim160$~K ($\sim J$), and their slow fluctuations freeze below $\sim50$~K \cite{Wang2022}.  That is, the crystal structure of Zn$_{0.95}$ is {\it locally} distorted without breaking the overall ideal kagome rotational symmetry of the lattice.  The crystal structure of Cu$_4$ is also distorted at low temperatures, as evidenced by the development of the superlattice structural peak in neutron scattering experiments \cite{Smaha2020}, prior to the drastic growth of $\chi_\text{bulk}$ below $\sim100$~K.  It is possible that these paramagnetic domains in Zn$_{0.95}$ are stabilized and grow in size owing to the local lattice distortion. 

Turning our attention to the $1/T_1$ results in Fig.~8, $1/T_{1}^\text{para} \simeq 100$~s$^{-1}$ observed for the paramagnetic peak of Zn$_{0.95}$ is nearly temperature independent, and comparable to the high temperature asymptotic value of $1/T_1$ observed for Cu$_4$.  In contrast, Cu$_4$ exhibits a continuous enhancement of $1/T_1$ with decreasing temperature due to the growth of short range order toward the long range antiferromagnetic order at $T_{N} = 15$~K.  According to Eq.(1), the local dynamical spin susceptibility in the vicinity of the interlayer Cu$^{2+}$ defects of Zn$_{0.95}$ therefore grows slowly as $\chi'' \sim 1/T$.  This means that gapless low energy spin excitations exist locally in the vicinity of interlayer Cu$^{2+}$ defects.  We emphasize, however, that these gapless spin excitations in Zn$_{0.95}$ are driven by the presence of interlayer defects and their influence on local magnetic and structural environments of the kagome planes, and  should not be confused with the intrinsic gapless spin excitations expected for certain theoretical models of pristine kagome spin liquids \cite{Ran2007,Liao2017}.  

In this context, it is worth recalling that some gapless spin liquid models predict a power law behavior in the temperature dependence of $1/T_1$ \cite{Ran2007}.  The recent $^{17}$O $1/T_1$ results \cite{Khunita2020} as well as our initial $^{63}$Cu $1/T_1$ results \cite{Imai2008} for herbertsmithite, both deduced from the empirical analysis of $M(t)$ due to extremely large distributions of $1/T_1$, appeared to confirm such expectations of a power law behavior.  However, our recent ILTT$_1$ analysis of $1/T_1$ observed at $^{63}$Cu sites showed that such an apparent power law temperature dependence previously proposed for herbertsmithite is purely fictitious \cite{Wang2021}.  Cu sites exhibit either gapped spin singlet behavior with activation type temperature dependence of $1/T_1$ or paramagnetic behavior with large, nearly temperature independent $1/T_{1} \sim 10^{3}$~s$^{-1}$.  The latter is qualitatively similar to $1/T_{1}^\text{para}$ observed here at paramagnetic $^{19}$F sites.   This suggests that the paramagnetic $^{63}$Cu sites observed for both Zn$_{0.95}$ and herbertsmithite \cite{Wang2021} may be also located in the same spin polarized domains observed here.

The absence of the signature of critical slowing down or spin freezing at 15~K for $1/T_{1}^\text{para}$ implies that the dimensions of the  spin polarized domain remain finite.   Assuming that these domains are centered around 5\% interlayer defect spins, $F_\text{para} \sim 60$\% at 2~K implies that the domains grow in space but span only a few triangles from defects, as schematically shown in Fig.~1c.  The significant fraction $F_\text{para} \sim 60$\% at 2~K with strongly enhanced  $1/T_{1}^\text{para}$ also implies that, as far as the low energy spin excitations probed by NMR are concerned, one can expect to find the intrinsic kagome spin liquid behavior associated with spin singlet formation only in $1-F_\text{para}\sim40$\% or less of the sample volume.  In Fig.~10, we show the temperature dependence of $1-F_\text{para}$ together with the upper bound of the spin singlet fraction $F_\text{singlet}$ as determined by $^{63}$Cu NQR measurements \cite{Wang2021}.  Indeed the upper bound $F_\text{singlet}$ in Zn$_{0.95}$ is  limited by the fraction of non-paramagnetic domains $1-F_\text{para}$ without spin polarization induced in the vicinity of interlayer Cu$^{2+}$ defects.  We recall that $F_\text{singlet}$ is an upper bound \cite{Wang2021} and could easily be overestimated by a factor of $\sim2$ due to difficulties in accounting for the transverse relaxation effects, but the magnitude of $F_\text{singlet}$ is still significant in comparison to $1-F_\text{para}\sim40$\%.

In conclusion, we have found two distinct magnetic signatures in the quantum spin liquid candidate Zn-barlowite, Zn$_{0.95}$. One signature arises from domains that are nearest to interlayer Cu$^{2+}$ impurities, and the local susceptibility and spin-lattice relaxation rate are similar to the paramagnetic state above $T_N$ of the antiferromagnetic parent compound barlowite, Cu$_{4}$. The other signature arises from regions far from the impurities and therefore more closely reflects the intrinsic behavior of the kagome spins, where spin singlets gradually emerge with spatially inhomogeneous gaps~\cite{Wang2021}.  Finally, it may be also interesting to apply the one- and two-dimensional NMR techniques based on ILT to other kagome spin liquid and related materials, which are known to exhibit distributions in their NMR properties~\cite{Kermarrec2014,Klanjsek2017,Lu:2022aa}.  Moreover, our approach might yield fresh insight into the distributions in local magnetic properties in unrelated disordered materials, such as diluted magnets~\cite{Itoh1986} and heavy Fermions~\cite{JohnstonPRL2005}.\\

\noindent{\bf METHODS}\\

\indent We synthesized the deuterated (D = $^{2}$H) powder samples of ZnCu$_{3}$(OD)$_{6}$FBr  and  Cu$_{4}$(OD)FBr  based on the procedures described in detail in \cite{Smaha2020}.  We confirmed the sample quality based on powder X-ray diffraction measurements.  We conducted the spin echo $^{19}$F NMR measurements using standard pulsed NMR spectrometers.  We conducted most of the $1/T_1$ measurements by applying saturation comb pulses prior to the spin echo sequence, and confirmed that the results are the same if we use an inversion pulse instead.  The pulse separation time $\tau$ between the 90$^{\circ}$ and 180$^{\circ}$ pulses were 10 $\mu$s to ensure that we do not overlook any signals with faster transverse relaxation time.  We carried out the inverse Laplace transform of the recovery curve $M(t)$ based on Tikhonov regularization using a fixed regularization parameter $\alpha =1$ rather than optimizing at different temperatures and frequencies.  This ensured that the effective resolution of ILT remains unchanged between different sets of measurements. \\

\noindent{\bf DATA AVAILABILITY}\\

The data sets generated and/or analyzed during the current study are available from the corresponding author on reasonable request.\\

\noindent{\bf ACKNOWLEDGMENTS}\\

The work at McMaster was supported by NSERC.  P.M.S. was supported by the Rice University Consortium for Processes in Porous Media.  The work at Stanford and SLAC (sample synthesis and characterization) was supported by the U.S. Department of Energy (DOE), Office of Science, Basic Energy Sciences, Materials Sciences and Engineering Division, under contract no. DE-AC02-76SF00515.  R.W.S. was supported by a NSF Graduate Research Fellowship (DGE-1656518). \\

\noindent{\bf AUTHOR CONTRIBUTIONS}\\

WY and J. Wang made equal contributions.  TI and YSL planned the project.  RWS, J. Wen, and YSL grew and characterized the samples.  WY, J. Wang, PMS, and TI carried out NMR measurements and analyzed the data.  TI and J. Wang wrote the manuscript with the input from all authors.\\

\noindent{\bf COMPETING INTERESTS}\\

The Authors declare no Competing Financial or Non-Financial Interests.\\

\newpage


\newpage

%\nocite{*}

%\bibliography{Yuan_npj_2022_v3f_arXiv}% Produces the bibliography via BibTeX.

%\newpage

\end{document}